\documentstyle[11pt,aaspp4,flushrt,tighten,epsf]{article}

\def\mathnew{\mathsurround=0pt}
\font\trm=cmr10

\def\etal{{\it et al.\ }}
\def\ie{{\it i.e.\ }}
\def\cf{{\it cf.\ }}
\def\eg{{\it e.g.\ }}
\def\kms{\ {\rm km\,s^{-1}}}
\def\hmpc{\ h^{-1}{\rm Mpc}}

\def\ds{D_{\rm S}}
\def\dl{D_{\rm L}}
\def\dls{D_{\rm LS}}
\def\amin{A_{\rm min}}
\def\sigcr{\sigma_{\scriptscriptstyle\rm C-R}}
\def\simov#1#2{\lower .5pt\vbox{\baselineskip0pt \lineskip-.5pt
       \ialign{$\mathnew#1\hfil##\hfil$\crcr#2\crcr\sim\crcr}}}

\def\today{\ifcase\month\or
  January\or February\or March\or April\or May\or June\or
  July\or August\or September\or October\or November\or
December\fi
  \space\number\day, \number\year}

\eqsecnum
\begin{document}

\title{Gravitational lensing of type Ia supernovae by galaxy clusters}

\author{%
  Tsafrir S. Kolatt\altaffilmark{1}
    \affil{The Physics Department and Lick Observatory, University of
    California, Santa Cruz, CA 95064, U.S.A.}
  \and
  Matthias Bartelmann\altaffilmark{2}
    \affil{Max-Planck-Institut f\"ur Astrophysik, P.O.\ Box 1523,
    D--85740 Garching, Germany}
\altaffiltext{1}{email:tsafrir@physics.ucsc.edu}
\altaffiltext{2}{email:mbartelmann@mpa-garching.mpg.de}}

\begin{abstract}
We propose a method to remove the mass sheet degeneracy that arises
when the mass of galaxy clusters is inferred from gravitational shear.
The method utilizes high-redshift standard candles that undergo weak
lensing. Natural candidates for such standard candles are type Ia
supernovae (SN Ia).

When corrected with the light-curve shape (LCS), the peak magnitude of
SN Ia provides a standard candle with an uncertainty in apparent
magnitude of $\Delta m\simeq 0.1-0.2$. Gravitational magnification of
a background SN Ia by an intervening cluster would cause a mismatch
between the observed SN Ia peak magnitude compared to that expected
from its LCS and redshift. The average detection rate for SN Ia with a
significant mismatch of $\ge2\Delta m$ behind a cluster at
$z\simeq0.05-0.15$ is about 
$1-2$ supernovae per cluster per year at $J,I,R\lesssim25-26$.

Since SNe are point-like sources
for a limited period, they can experience significant
microlensing by MACHOs in the intracluster medium. Microlensing events
caused by MACHOs of $\sim10^{-4}\,{\rm M}_\odot$ are expected to have
time scales similar to that of the SN light curve.
Both the magnification curve by a MACHO and the light curve of a SN Ia
have characteristic shapes that allow to separate them.
Microlensing events due to MACHOs of smaller mass can unambiguously be
identified in the SN light curve if the latter is continuously
monitored. The average number of identifiable microlensing events per
nearby cluster ($z\lesssim0.05$) per year is 
$\sim 0.02 \,(f/0.01)$, where $f$ is the fraction of the cluster mass in MACHOs 
of masses $10^{-7} < {\rm M_{\rm macho}}/{\rm M}_\odot < 10^{-4}$. 
\end{abstract}

\section{Introduction}
\label{sec:intro}

The mass sheet degeneracy (Falco, Gorenstein, \& Shapiro 1985;
Schneider \& Seitz 1995) constitutes one of the fundamental
uncertainties in attempts at reconstructing galaxy cluster masses from
the gravitational distortion of resolved background sources.  
It arises because the observed gravitational shear field is
insensitive to magnification.  A few methods were proposed to break
the degeneracy by directly measuring either the magnification or the
(scaled) surface mass density. Broadhurst, Taylor, \& Peacock (1995)
suggested to compare the redshift and/or the magnitude distribution of
the background sources in cluster fields with those measured in empty
fields.  Kaiser (1995) and Kneib \etal (1995) showed that inferring
the mean source redshift at a given apparent magnitude can help to
break the degeneracy. Bartelmann \& Narayan (1995) proposed and tested
an algorithm that exploits the size distribution of the background
sources and the conservation of surface brightness under gravitational
lensing.  An alternative cluster reconstruction technique suggested by
Bartelmann \etal (1996) breaks the mass sheet degeneracy by
simultaneously taking sizes {\em and\/} shapes of background galaxies
into account.  All these methods rest upon statistical information on
intrinsic background galaxy sizes. They require accurate measurements
of sizes and magnitudes or surface brightnesses of lensed galaxies. In
principle, a single sufficiently precise measurement of the
magnification of a standard candle by a cluster can lift the mass
sheet degeneracy as well.

Suppose a standard candle is observed behind a lens, and its redshift
is measured. Assuming a cosmological model, the luminosity distance is
known. The apparent magnitude expected from its luminosity can then be
compared to the observed one, and any discrepancy between the two can
be attributed to the magnification by the foreground lens. Ideal
standard candles are rare in cosmology. Nevertheless, there is
accumulating strong evidence that SN Ia may well serve as approximate
standard candles, especially if one takes the dependence of their peak
magnitude on the light curve shape into account (Phillips 1993; Riess,
Press, \& Kirshner 1995 (RPK1); Riess, Press, \& Kirshner 1996 (RPK2);
Hamuy \etal 1996a).  Moreover, SN Ia are ubiquitous, point-like
sources, their distinct spectral lines allow for an accurate redshift
determination, and as of yet they show no cosmological evolution.
Though rare in frequency, each of their aforementioned characteristics
makes SN Ia excellent probes for the integrated matter density along
their lines-of-sight, and the composition of the matter (\ie its
graininess, occurrence of MACHOs etc.).  Weak lensing by large-scale
structure was shown to not significantly affect the SN magnitude
(Frieman 1996; Wambsganss \etal 1996), but this does not apply to
galaxy clusters.

Similar arguments led Kovner \& Paczy\'nski (1988) to propose a
possible identification of SNe in giant arcs as a tool to establish
the time delay between different images. Giant arcs are less appealing
than weak lensing by clusters because they are much more rare, and
sensitive to the details of the density profile close to the cluster
core.

Soucail \& Fort (1991) suggested the combination of giant arcs and a
much less accurate ``standard candle'', viz.\ the Tully-Fisher
relation, to estimate the Hubble constant.  This method suffers from
several difficulties, \eg a 
possible cosmological evolution of the Tully-Fisher relation, and the
problematic determination of the inclination angle of the lensed
galaxy.

In the rest of this paper, we present the basic considerations one
has to take into account in order to employ SN Ia for these
purposes. In \S\ref{sec:sn_cl} we review the necessary facts about the
SN population and the cluster model we use. In \S\ref{sec:esti} we
estimate the number of expected SN in the background of a cluster,
whose luminosity--distance mismatch is large enough to allow for the
removal of the mass sheet degeneracy. In \S\ref{sec:machos} we discuss
the use of a lensed SN Ia as a probe for MACHOs in the intracluster
medium. We conclude with our results in \S\ref{sec:conc}.

\section{Supernovae and clusters}
\label{sec:sn_cl}

\subsection{Supernovae Type Ia}
\label{subsec:sn}

\begin{figure}[t!]
\hskip0.5truecm {\epsfxsize=2.7 in \epsfbox{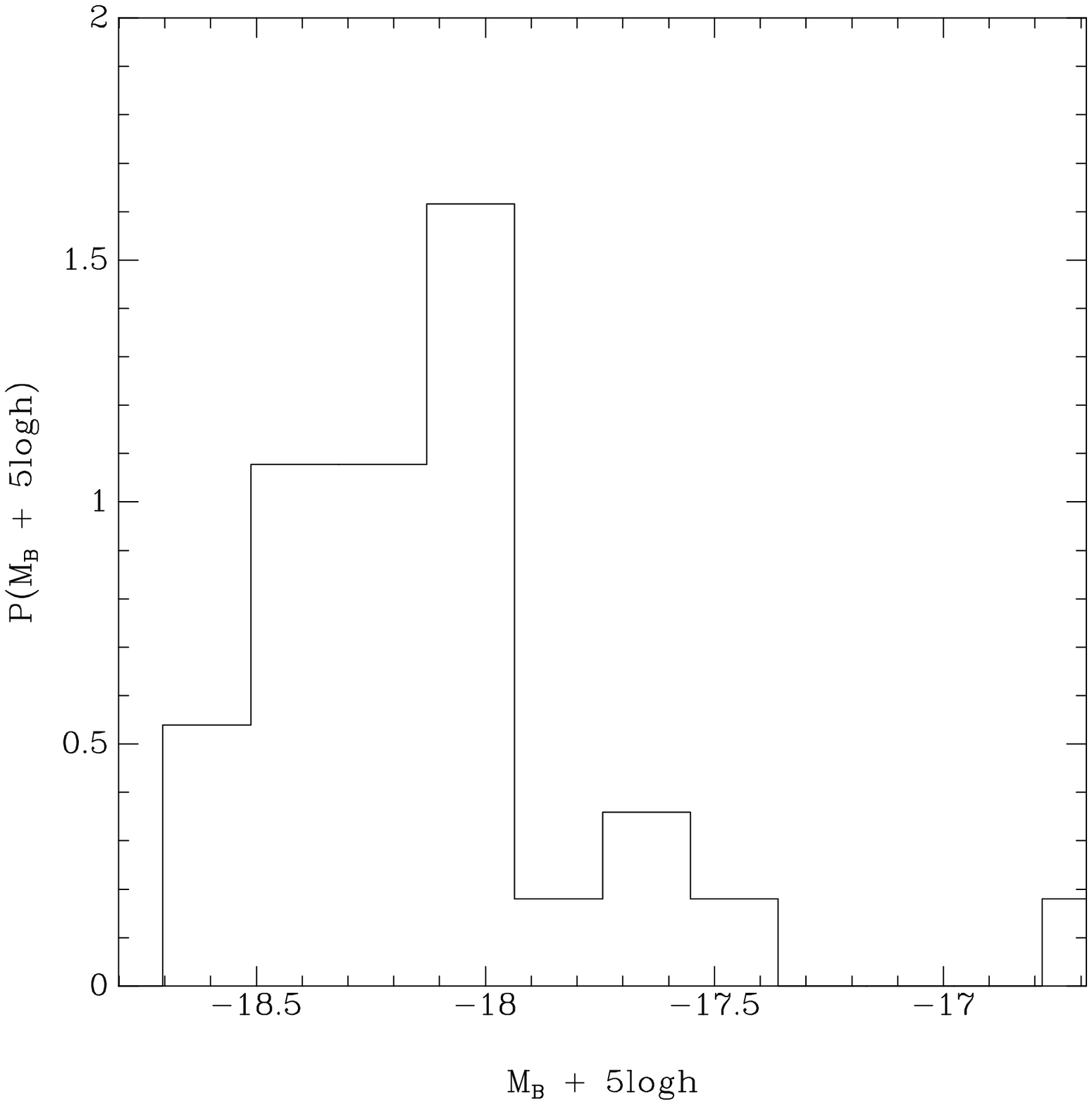}}
\hskip2truecm {\epsfxsize=2.7 in \epsfbox{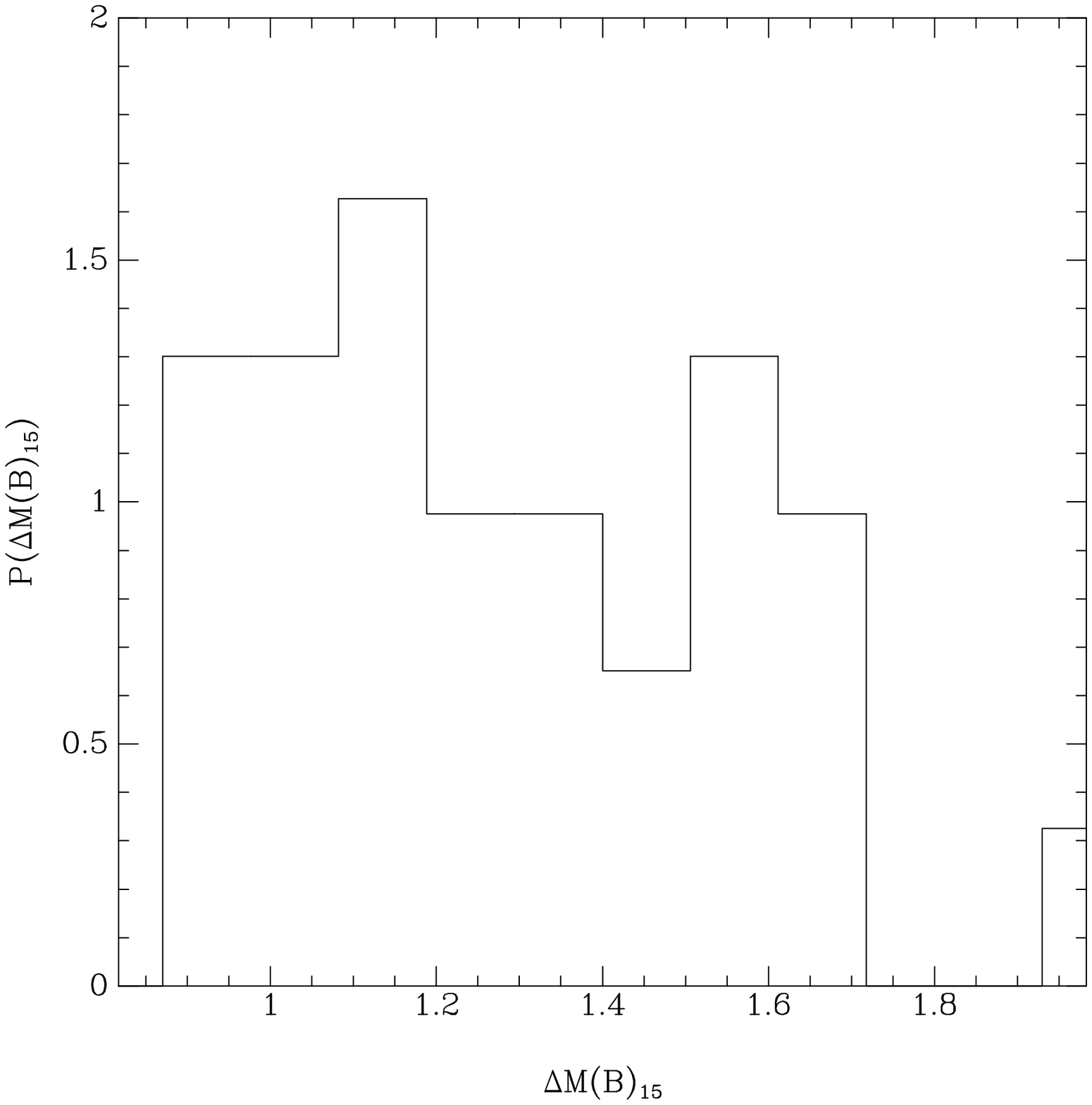}}
\vskip-0.5truecm
\caption{\baselineskip 0.4cm{\trm The frequency distribution of the
  absolute B magnitude from the Hamuy \etal (1996a,b) sample of 29 SNe
  (left), and the distribution of the magnitude difference $15$
  (rest-frame) days past the peak for the same sample (right).}}
\label{fig:dis_sn}
\end{figure}

The distribution of absolute peak magnitudes of SN Ia is intrinsically
narrow (see Branch \& Tammann 1992 for a review).  From the Hamuy
\etal (1996a,b) sample, we find the average absolute magnitude and its
standard deviation to be $(-18.12,0.38)$, $(-18.19,0.26)$, $(-17.97,
0.19)$, for $\langle M_{\rm SN} \rangle + 5\log h$ in the B, V, and I
bands, respectively\footnote{Hereafter, $M_{\rm SN}$ is the absolute
blue peak magnitude.}.  The quoted standard deviation is larger than
the intrinsic one, since Hamuy \etal did not correct for reddening due
to dust in the host galaxies. Figure \ref{fig:dis_sn}a shows the
absolute-magnitude distribution of the sample. These results are in
accord with Branch \& Miller (1993) [($-18.2+5\log h,0.36)$ in B], and
Vaughan et al.\ (1995) [($-18.28+5\log h, 0.31)$ for a
sub-sample]. The latter two results were obtained by cross-comparison
with other distance indicators, that are only available for relatively
nearby galaxies.  The results may therefore suffer from the intrinsic
scatter in these other distance indicators. The Hamuy \etal (1996a)
sample includes distant galaxies (up to $z=0.1$) whose redshift
provides a good enough distance estimate.

A substantial decrease in the scatter of the peak luminosity
distribution is achieved if a correction that is based on the shape of
the light curve (LCS) is applied in different bands 
(Phillips 1993; RPK1, RPK2, Hamuy \etal 1996a).
This can also account for reddening due to dust in the host galaxy,
the Milky Way,
and possibly in the foreground cluster.
After the correction, SN Ia can be regarded as standard
candles up to a small dispersion of $\Delta m \simeq 0.1-0.2$.

A parameter that is traditionally taken to govern the LCS is the
decline in B magnitude 15 rest-frame days after maximum light, $\Delta
M_{15}$. According to the Hamuy \etal (1996a) data, its distribution
has a mean and standard deviation of $(1.28, 0.28)$, respectively.
Figure \ref{fig:dis_sn}b shows the distribution of the rest-frame
$\Delta M_{15}$ from the 29 Hamuy \etal (1996a) SNe.  The small
dispersion in $\Delta M_{15}$ is a reflection of similar timescales
for the rise and decline of the light curves of different SN.  In
\S\ref{subsec:rate} we shall use a rise time of 5 rest-frame days.
This corresponds to a magnitude difference of about $\Delta
M_{-5}=\vert M_{\rm SN}-M_{\rm SN-5}\vert \sim 0.1-0.3$.

The specific rate of SNe Ia, ${\cal S}_{\rm SN}$, (measured in SNu: SN
per $10^{10}L_{B\odot}$ per century) shows a rise with redshift.  The
local specific rate, ${\cal S}_{\rm SN}(z=0)$, is estimated to lie
between $0.2h^2$ and $0.7h^2$ SNu, depending on the host galaxy type
(Cappellaro \etal 1993a, 1993b; Turatto, Cappellaro, \& Benetti 1994;
van den Bergh \& McClure 1994; Muller \etal 1992). The average rate at
$z=0.4$ is estimated to be $\simeq0.82 h^2$ SNu (Pain \etal 1996). The
rate evolution may depend on galaxy evolution, the stellar mass
function, different composition of late and early type galaxies at
different redshifts, and a different fraction of massive blue stars
that undergo a SN Ia phase.

Even though the evidence for a rise with redshift is still
inconclusive, all of these dependencies lead to an increment of the
average specific rate of SN Ia with redshift.  Hamuy \etal (1995)
suggest that galaxies with younger stellar population produce the most
luminous SNe. Branch, Romanishin, \& Baron (1996) argue that SNe Ia
preferentially reside in late type spirals. It is hence natural to
expect a higher average specific rate of SN Ia at higher redshifts.

\subsection{The Cluster Model}
\label{subsec:cluster}

All theoretical and observationally motivated models for cluster
density profiles agree on its parametric shape ($\propto r^{-2}$) at
intermediate radii ($\sim 0.1-1.5\hmpc$) between its core and
outskirts (\eg Gunn \& Gott 1972; Filmore \& Goldreich 1984;
Bertschinger 1985; Hoffman 1988). 
Mass profiles obtained from
weak lensing corroborate this parametric form 
for a few specific cluster examples (\eg Tyson \& Fischer 1995).
N-body simulations show a similar
behavior (\eg Navarro, Frenk, \& White 1996).  We hence use a singular
isothermal sphere (SIS) as our cluster model, \ie a density profile of
$\rho(r) = (\sigma_v/c)^2/(2\pi G) r^{-2}$, where $\sigma_v$ is the
one-dimensional velocity dispersion. Cluster cores, if they exist, are
of size $\lesssim100\,h^{-1}$kpc, beyond which the cluster profile
resembles the SIS profile.
Note that this is the scale where weak lensing
starts to serve as a tool for deriving the cluster density profile,
and thus where the mass-sheet degeneracy comes in.
Since most SN Ia lensed by clusters are
expected to be found at radii larger than the core, the approximation
of a SIS profile is good enough for our purposes.  At the other end,
we are limited by the minimal magnification for a mismatch
identification. This requirement sets the outer radius of interest.
This radius is always smaller than the scale where the asymptotic
behavior of the cluster density profile differs from SIS for different
models.

For a SIS, the magnification of an image at angular distance 
$\theta$ from the cluster center is given by 
\begin{equation}
\label{eq:a_sis}
  A(\theta) = \left[
  1 - {\theta_{\rm E} \over \theta }
  \right]^{-1}\;;\quad
  \theta_{\rm E} = {\dls \over \ds}\;4 \pi
  \left({\sigma_v \over c}\right)^2 \,.
\end{equation}
$\ds$ and $\dls$ are the angular diameter distances to the source, and
between the lens and the source, respectively.  $\theta_{\rm E}$ is
the Einstein angle of the cluster.  For $\theta<2\theta_{\rm E}$, two
images appear on either side of the lens, the fainter one being at
$\theta<\theta_{\rm E}$, and the brighter one at $\theta>\theta_{\rm
E}$. We ignore the fainter image in the following, thus having
$\theta\ge\theta_{\rm E}$ and $A\ge1$ always.

The cross section for sources that are magnified by more than
$A\ge\amin\ge1$ by a SIS foreground lens is
\begin{equation}
\label{eq:cross}
  \sigma_{\rm SIS}(A) = {\pi \theta_{\rm E}^2 \over
  (A-1)^2}\,.
\end{equation}
Hence, magnifications $A\ge\amin$ are distributed according to
\begin{equation}
\label{eq:p_of_a}
  P(A) \propto -{ {\rm d} \sigma_{\rm SIS}(A) \over {\rm d} A }
  \;;\qquad
  P(A){\rm d} A = {2(\amin-1)^2 \over (A-1)^3 }{\rm d} A \,,
\end{equation}
where the second expression follows from the normalization $
\int_{\amin}^\infty P(A){\rm d}A = 1 \,.  $

\section{Estimation of the observed SN rate}
\label{sec:esti}

We model the quality of the SN to serve as standard candles in terms
of the natural width, $\Delta m$, of the LCS--magnitude--distance
relation.

The magnification by a cluster boosts the apparent magnitude by
$-2.5\log_{10}A(\theta)$.
If the boost is large enough, a mismatch between the SN apparent
magnitude (corrected by the LCS) and its luminosity distance will show
up.

The luminosity distance is calculated from the SN redshift, given the
cosmological parameters ($\Omega_{\rm m}$, $\Omega_\Lambda$,
($\Omega_k = 1 - \Omega_{\rm m} - \Omega_\Lambda$) 
and $H_0$).  The closer the SN is to the cluster center, the larger 
is the magnification $A$, and the more pronounced is the discrepancy.
The boost allows to observe sources that without the presence of the
lens would fall below the detection flux limit.  A competing effect of
the lens is the reduction of the surface number density of background
sources by $A(\theta)^{-1}$.

The bigger the cluster appears on the sky, the more background
galaxies it encompasses, and thus the 
more potential SNe.  This geometrical effect must be combined with the
competing $\dls/\ds$ term in the magnification (eq. \ref{eq:a_sis}).

We assume all SNe reside in galaxies. We do not distinguish between
early and late type galaxies, as SN Ia occur in both, though there is
an indication 
that SNe are more frequent in late-type galaxies (Cappellaro \etal
1997). The relevant function one needs to know is thus $n(z)$, the
number density of galaxies at redshift $z$ , and their luminosity
distribution.  It is also essential to know what the average number of
SNe is for these galaxies at a given redshift.  If the rest-frame blue
luminosity function of galaxies at the relevant redshift is known, we
may adopt a value for the specific SN rate, ${\cal S}_{\rm SN}$, and
obtain the SN Ia average rate per proper time unit per comoving volume
unit from
\begin{equation}
\label{eq:sn_in_vol}
    S_{\rm SN}(z) = {\cal S}_{\rm SN}(z) \,
    \int_{-\infty}^\infty {\rm d}L\, L\, P(L,z)\;,
\end{equation}
with $P(L,z)$ the Schechter luminosity function at 
redshift $z$ 
(\cf \S\ref{sub_sec:ng_of_z}).  For this calculation it is essential
to have an estimate for $P(L,z)$, the normalization and the 
parameters of the Schechter function at each redshift, or its integral
-- the number density of galaxies at each redshift.

\subsection{Galaxy number density at different redshifts}
\label{sub_sec:ng_of_z}

The best estimate to date for the number density of galaxies as
function of redshift at $z>0.3$ comes from the Canada-France redshift
survey (CFRS). Lilly \etal (1995) estimated the Schechter function
parameters in the rest-frame $B$ band in three redshift
bins\footnote{We translated the CFRS $B_{AB}$ to $B$ to comply with
the notation in the rest of this paper.}.  At lower redshift, we
adopt the estimates for these parameters from the APM survey (Loveday
\etal 1992) and the AUTOFIB survey (Ellis \etal 1996). In order to get
the overall galaxy density in a given redshift bin, we integrate the
Schechter function with the adopted parameters over the range
$(-\infty,M_{\rm lim}]$.  $M_{\rm lim}$ is the minimum absolute
magnitude that prevents divergence of the integral.

\begin{figure}[t!]
\vskip1.0truecm
\hskip4truecm{\epsfxsize=3.3 in \epsfbox{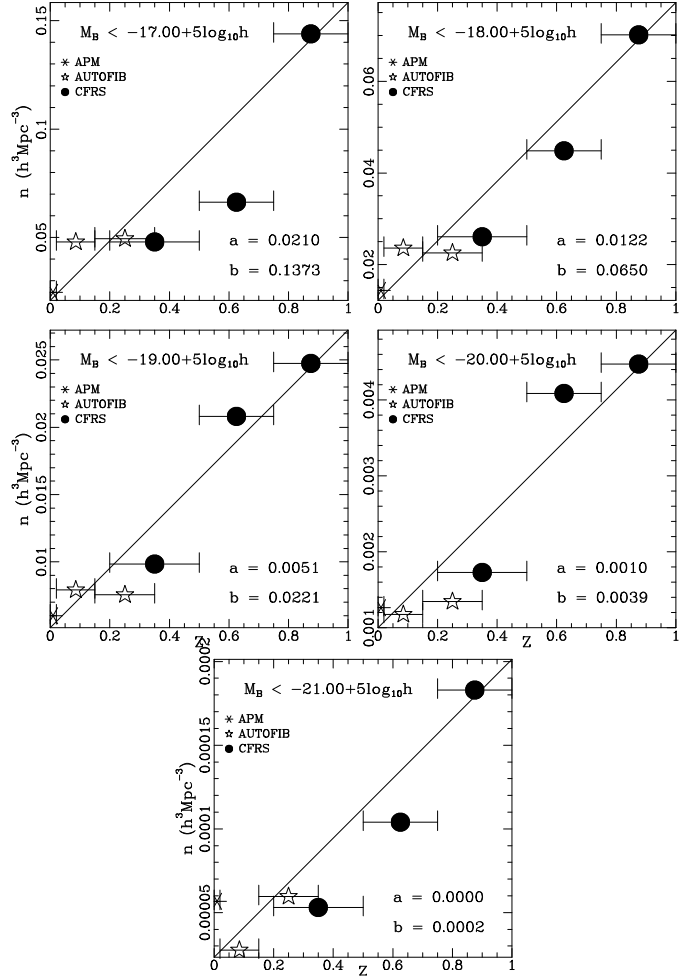}}
\vskip0.8truecm
\caption{\baselineskip 0.4cm{\trm 
  Spatial number density of galaxies with
  $M_B$ (rest-frame) brighter than $-17$ $(+5\log_{10}h)$, $-18$,
  $-19$, $-20$, and $-21$ absolute magnitude, as function of redshift.
  All points are obtained by integrating 
  over the Schechter luminosity function with appropriate parameters
  as quoted by the different authors. The diagonals are linear
  approximations (not fits) to the functions, and their parameters are
  displayed ($n(z) = a+bz$).}}
\label{fig:den_gal}
\end{figure}

Figure \ref{fig:den_gal} shows the number density for five different
values of $M_{\rm lim}$ as a function of redshift. For the CFRS, we
took the parameters of their ``best'' estimate with $q_0=0.5$. The
points represent the center of the bin (not volume weighted), and the
diagonal line is an approximate behavior of a linear function guided
by all the surveys' points\footnote{We did not perform a least square
linear fit because Lilly \etal warn the reader not to take their
fitting parameters beyond their fitting range, and because the errors
in different surveys do not necessarily have a common reference
ground. We tried to always {\em underestimate\/} the function
slope.}.

The highest redshift bin of the CFRS ends at $z=1$. For the current
analysis we need to extrapolate $n(z)$ beyond that limit. We need to
know how far one can safely extrapolate the approximation, and to what
extent the functional form and the derived parameters are adequate.

A good verification for 
estimates of the galaxy number density as a function of redshift is
the calculated surface number density of galaxies in a given apparent
magnitude limit. Observed values for the surface number density range
between $50-100$ arcmin.$^{-2}$ for $B\sim25-27.5$, corresponding to
$R\sim24-26.5$ (Tyson 1988; Lilly, Cowie, \& Gardner 1991; Metcalfe
\etal 1991, 1995), and as high as $\sim200$ arcmin.$^{-2}$ for
$R\sim27$ (Smail \etal 1995).  If we consider all galaxies with
$M<-17+5\log h\, (=M_{\rm lim})$, and apply an apparent magnitude
limit of $m<26$, and use the relations given in Fig.~\ref{fig:den_gal}
($n(z)=a+bz$), we find that integration out to $z_{\rm S}\simeq2.0$
yields surface densities of $\sim60-70$ arcmin.$^{-2}$.  Integration
out to $z=1.5$ yields $45$ arcmin.$^{-2}$. Note that we did not model
the K-correction, and assigned only rest-frame B magnitudes to the
galaxies. The limit we quote is thus roughly equivalent to an R or I
magnitude limit.

Since the integration we performed over the functions shown in 
Fig.~\ref{fig:den_gal} along redshift yields a similar surface number
density as observed, we may safely adopt the form $n(z)=a+bz$ as our
estimate for the redshift dependence of galaxy number density.  We
extrapolate the functional form and parameters from the observational
data to $z=2.0$ in order to achieve the right surface number density,
but we do not attempt to extrapolate any further, even though it may
increase the resultant rate of SN Ia detection.  The parameters ($a$
and $b$ of the last expression) for different cutoffs in rest-frame
absolute B magnitude appear in fig. \ref{fig:den_gal}. Recall we need
the galaxy number density only in order to know the size of the
reservoir from which observable SNe are taken, and to estimate the
upper limit of the redshift.  The magnitude limit of the proposed SN
Ia survey is irrelevant to the host galaxy luminosity.

\subsection{The rate calculation}
\label{subsec:rate}

\begin{figure}[!t]
\plotfiddle{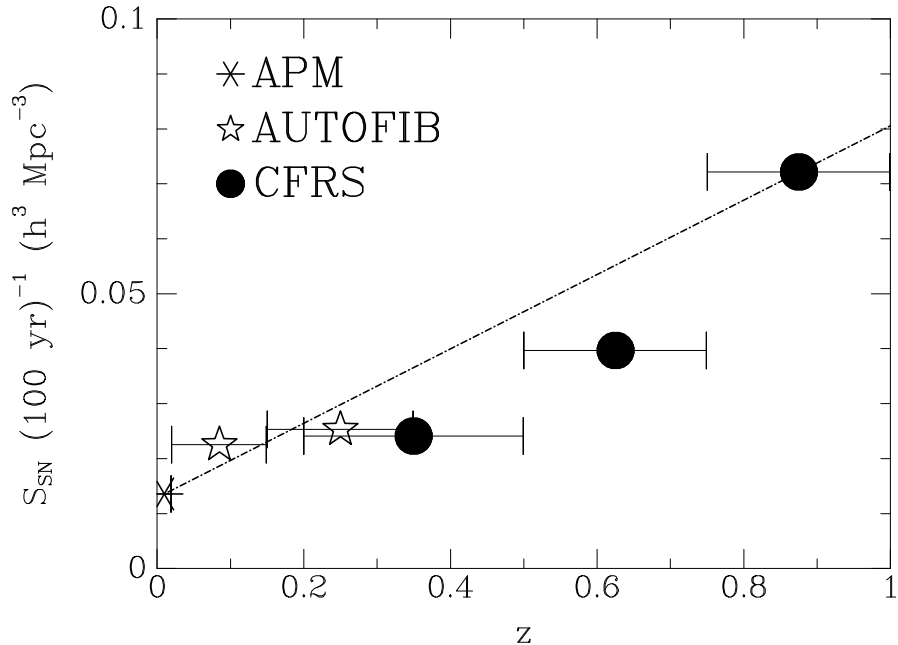}{3truecm}{270}{50}{50}{-220}{140}
\vskip4.5truecm
\caption{\baselineskip 0.4cm{\trm The calculated rate of SN per
  century per unit of comoving volume as function of redshift
  (eq. \ref{eq:sn_in_vol}).  The specific rate is taken to be ${\cal
  S}_{\rm SN}(z=0.4)=0.82h^2$ SNu.  Symbols, Schechter luminosity
  function parameters, and redshift bins are identical to those used
  in fig.~\ref{fig:den_gal}. The straight line is our underestimating
  model (not a fit, see text) for the rate evolution.}}
\label{fig:sn_in_vol}
\end{figure}

Using the same parameters for the Schechter luminosity function, we
calculate the SN rate per century per unit of comoving volume, $S_{\rm
SN}$, at different redshifts (eq. \ref{eq:sn_in_vol}).  Figure
\ref{fig:sn_in_vol} shows this rate.  We choose our base-line specific
rate to be ${\cal S}_{\rm SN}=0.82h^2$ SNu from the $z=0.4$ estimate
(Pain \etal 1996).  We expect most of the host galaxies to lie beyond
this redshift. We ignore a possible increase of this base-line with
redshift.  The change of $S_{\rm SN}$ in fig. \ref{fig:sn_in_vol} is
therefore entirely due to the change in the Schechter function
parameters.  We model this change crudely by a linear function $S_{\rm
SN} = 0.0136+0.0670z$ (100 yr)$^{-1}$ $(\hmpc)^{-3}$; the straight
line in fig. \ref{fig:sn_in_vol} describes this curve. This line is
not a least-square fit, for reasons mentioned earlier. We anchored the
line at the $z=0$ value since the Schechter function parameters are
best determined in the local universe.  It seems that a quadratic
curve will fit the estimated data better; however, extrapolation of
such a curve would not be a conservative choice for $z>1$.  Moreover,
fig. \ref{fig:den_gal} suggests a linear behavior of $n(z)$, the most
important quantity that governs eq. \ref{eq:sn_in_vol}.  These
considerations lead us to choose the linear approximation for $S_{\rm
SN}(z)$ as it appears in the plot.

In order to get the expression for the expected rate of SNe behind a
lens with a $j\sigma$ magnitude--distance mismatch, we use the cross
section and magnification probability
(eqs. \ref{eq:cross}--\ref{eq:p_of_a}). The minimal magnification
needed in order to establish the luminosity--distance discrepancy is
$\amin$. For a $j\sigma$ discrepancy, $\amin=10^{0.4\,j\Delta m}$.
The expected rate is given by
\begin{eqnarray}
  {\cal R} &=& {1 \over 100}\,
  \int_{z_{\rm L}}^\infty {\rm d}z {{\rm d}R(z) \over {\rm d}z} R^2(z) \,
  {S_{\rm SN}(z) \over (1+z)} \, \sigma_{\rm sis}(\amin)\,
  \int_{\amin}^\infty {\rm d} A\,P(A) \nonumber\\
  &\times&
  \int_{-\infty}^\infty {\rm d}M_{\rm SN}\,P(M_{\rm SN})\,
  \Theta \left[ m_{\rm lim} -(M_{\rm SN} + \Delta M_{\rm th} + \mu(d^l_S) -
  2.5\log_{10}(A) ) \right] \, {\rm yr}^{-1}\;,
\label{eq:rate}
\end{eqnarray}
with the radial component of a comoving volume element
($R^2{\rm d}R = (d^M)^2/(1+\Omega_k H_0^2(d^M)^2)^{1/2}\,{\rm d}\,(d^M)$
\cf Carroll, Press, \& Turner (1992) where $d^M(z)$ is the proper motion
distance.)
$\mu(d^l_S)$ is the distance modulus as a function of the luminosity
distance. We take the survey magnitude limit for SN identification (as
opposed to the follow-up) to be $m_{\rm lim}$. 
$\Theta$ is the Heaviside step function 
that imposes the magnitude threshold for the detection of a SN.
The threshold detection magnitude is expressed by $\Delta M_{\rm th}$,
\ie the difference between the peak magnitude and the identification
magnitude we require. The factor $(1+z)^{-1}$ accounts for the
cosmological time dilation.  As mentioned above, we neglect a possible
fainter image in our calculation.

A few comments have to be made regarding expression \ref{eq:rate}. 
To begin with, we have neglected the redshift dependence of ${\cal
S}_{\rm SN}$ ( which propagates to $S_{\rm SN}$), albeit we use a
relatively low redshift ($z=0.4$) for it.  This is in spite of the
expected (and observed) increase of the specific rate with
redshift. Even a mild increase reflects strongly in the expected rate.
Another comment on expression (\ref{eq:rate}) regards the
K-correction.  Strictly speaking, since we integrated over $P(M)$ for
rest-frame blue magnitude, a K-correction will shift SNe into a
different observed band for each redshift, and thus also to different
apparent magnitude (because of different zero points for different
magnitudes).  We did not include the K-correction for three reasons:

\begin{enumerate}

\item 
The K-correction for SNe of redshift higher than $z=0.5$ has 
not yet been calculated (Kim, Goobar, \& Perlmutter 1996).

\item 
The K-correction at $z\simeq1$ generally translates an 
apparent magnitude limit in B (say 26) to be smaller in R and I (say
25--25.5).
The signal-to-noise in R or I, though, decreases as well and therefore
requires longer integration time.

\item 
As Goobar \& Perlmutter (1995) pointed out, when one goes to higher
redshift, the optimal filter is selected accordingly. Perlmutter \etal
(1995) suggest that a more robust procedure (than K-correction) is to
compare the LC in the band equivalent to the rest-frame B band.

\end{enumerate}

In general, thus, all the apparent B magnitude limits we quote are
higher than the required 
magnitude limits if observations are carried out in the appropriate
bands (but recall that the signal-to-noise decreases in R and I).

We model the absolute magnitude distribution of the SNe Ia ($P(M_{\rm
SN})$ of eq. (\ref{eq:rate})) as a $\delta$-function in peak absolute
magnitude. As can be seen from figure \ref{fig:dis_sn}a, the observed
distribution is narrow, and slightly skewed. There is higher
probability to find brighter-than-average absolute magnitudes than
dimmer-than-average. By using a $\delta$-function at the average,
$P(M_{\rm SN})=\delta(\langle M_{\rm SN} \rangle - M_{\rm SN})$, we
underestimate the observed rate predicted by eq. (\ref{eq:rate}).

Choosing this distribution of $M_{\rm SN}$ simplifies eq.
(\ref{eq:rate}) to 
\begin{equation}
\label{eq:rate_d}
  {\cal R} = {1 \over 100}
  \int_{z_{\rm L}}^\infty {\rm d}z {{\rm d}R(z) \over {\rm d}z} R^2(z) \,
  {S_{\rm SN}(z) \over (1+z) } \times {\rm min}
  \left[\sigma_{\rm SIS}(A')\,,\sigma_{\rm SIS}(\amin)\right]\,,
\end{equation}
with
\begin{equation}
\label{eq:a_prime}
  A' = {\rm dex}\left[ 0.4\left( \langle M_{\rm SN} \rangle +
  \Delta M_{\rm th} + \mu(d^l_S)-m_{\rm lim} \right) \right ]\,.
\end{equation}

Already being equipped with $S_{\rm SN}$, the SN rate in
unit volume as a function of redshift (eq. \ref{eq:sn_in_vol}), we
calculate expression (\ref{eq:rate_d}) in order to find the expected
rate of observed SNe with luminosity--distance mismatch behind a
cluster.  We choose the cosmological parameters ($\Omega_{\rm m} = 1$,
$\Omega_\Lambda=0$, $h=1$).  We require a typical SN to be found about
5 (rest-frame) days prior to peak luminosity. This is equivalent to
$\Delta M_{\rm th}\equiv\Delta M_{-5}=\vert M_{\rm SN}-M_{\rm
SN-5}\vert$ of $\sim 0.1-0.3$ magnitudes. We assume a natural scatter
in the LCS-luminosity relation of $\Delta m \equiv \Delta M_{\rm SN} =
0.12$ (RPK2), and decide how many $\Delta M_{\rm SN}$ away from the
average relation we would like a SN to be detected in order to claim a
mismatch.
This value for $\Delta m$ may be too optimistic for existing LCS methods. 
In principle, however, it is achievable especially if outlier SN are
easyly identified and eliminated.

\begin{figure}[t!]
\vskip-2.0truecm
\hskip3.6truecm {\epsfxsize=3.7 in \epsfbox{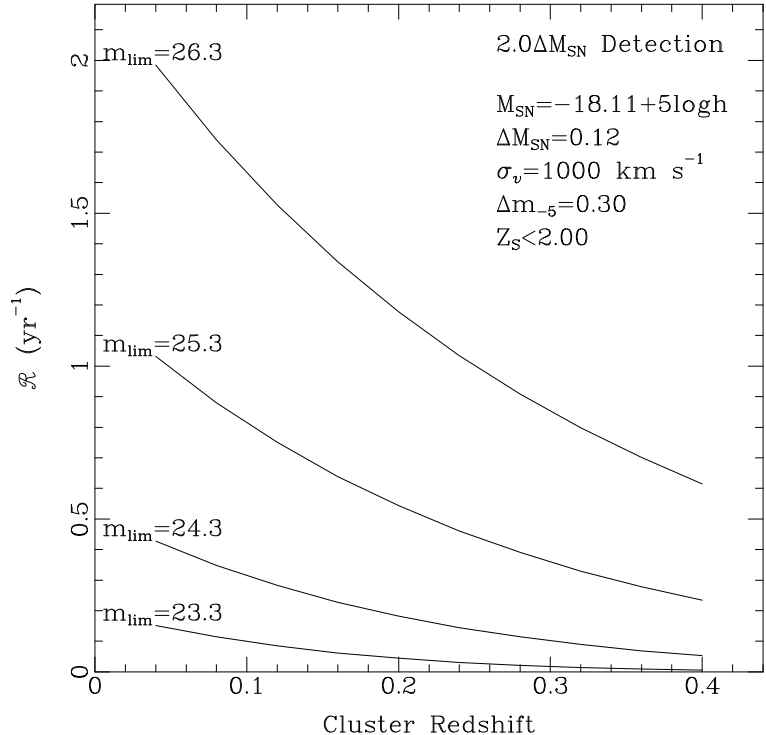}}
\vskip-1.0truecm
\caption{\baselineskip 0.4cm{\trm The expected detection rate of SNe
  Ia in the background of a lensing cluster as function of the lens
  redshift.  The imposed requirement is that the mismatch between the
  LCS-corrected peak luminosity and the distance be twice the scatter
  of the former.}}
\label{fig:rate}
\end{figure}

Figure \ref{fig:rate} shows the expected rate for 
{\em detectable\/} 
SNe behind a cluster at different redshifts.
We choose $\sigma_v=1000\kms$ for the cluster velocity
dispersion. 
The different curves correspond to different magnitude limits for the
SN detection. Recall that these limits are a direct translation from
the rest-frame B band to the actual observed bands, where one expects
the limit to be $\sim 1-1.5$ magnitudes lower (because of the
zero-point shift).  The integration over $z$ is carried out only up
to $z=2$ to achieve agreement with the observed surface number
density, and in order not to extrapolate the redshift dependence of
the galaxy number density too far. Since we did not include
K-correction in our calculation, we do not specify the band of the
magnitude limit. As was pointed out by Goobar \& Perlmutter (1995), it
is probably the I filter that is most suitable for $z\simeq 1$ SN
detection and the I,J  filters for SNe of $1<z<2$. Increasing the
magnitude limit beyond $26.3$ does not yield a higher rate because of
the cutoff in the source redshift.
The rate drops monotonically with lens redshift because the Einstein
angle of a SIS is almost independent of lens redshift as long as
$z_{\rm S}\gg z_{\rm L}$. The redshift dependence of the rate is thus
determined by the volume behind the lens that is spanned by the almost
constant angular cross section of the cluster. 

Figure \ref{fig:rate} indicates that if all LMC-like galaxies
($M_B<-17$ for $h=0.57$) host SNe with the same average specific rate
as observed at $z=0.4$ (the contribution of galaxies dimmer than $-17$
to $S_{\rm SN}$ is small), then we can expect to observe 1-2 SNe
behind each nearby cluster each year. Since most SNe will be at higher
redshift, the cosmic time dilation reduces the expected rate (\cf
eqs. \ref{eq:rate}, \ref{eq:rate_d}) but at the same time widens the
time window to catch the SN while its LC is rising.  The necessary
duration for the SN identification survey is therefore {\em not\/}
inversely proportional to the ratio between the local and high-$z$
expected rate.

\subsection{The degeneracy removal}
\label{subsec:removal}

If a mismatch of $\Delta m$ is found for a background SN, one has to
decide what is the permitted range of variation for the SN absolute
magnitude (recall the magnitude probability is asymmetric, \cf
fig. \ref{fig:dis_sn}a). For a permitted range of $k\sigma$, and a
$j\sigma$ mismatch detection, the lower and upper limits for the
magnification are ${\rm dex}[0.4(j-k) \Delta M_{\rm SN}] \le A \le
{\rm dex}[0.4(j+k) \Delta M_{\rm SN}]$.  These limits can immediately
be translated to limits on the cluster velocity dispersion,
\begin{equation}
  \left[ {\rm dex}[0.4(j-k) \Delta M_{\rm SN}] - 1 \right]
  {\theta \over C} \le \sigma_v^2 \le
  \left[ {\rm dex}[0.4(j+k) \Delta M_{\rm SN}] - 1\right]
  {\theta \over C} \,
\end{equation}
with $C=\theta_{\rm E}/\sigma_v^2$. The significance of these limits
is much increased when combined with other reconstruction methods (\eg
Kaiser \& Squires 1993; Squires \& Kaiser 1996; Schneider \& Seitz
1995; Seitz \& Schneider 1995, 1996, 1997; Bartelmann \etal 1996).

Being a point source, the SN probes a single line-of-sight that
intersects the cluster.  Inferring limits on the cluster mass from
limits on $A$ may bear two caveats; if the SN line-of-sight happens to
pass near a cluster galaxy, the galaxy adds to the magnification, and
the simple interpretation of the limits on $A$ in terms of
limits on the cluster mass is no longer valid. Barring dramatic
magnification bias, the two-dimensional filling factor of galactic
critical curves beyond 
$\theta_{\rm E}$ is very small, and therefore the galaxy contributions
can generally be ignored.

A similar contribution to the magnification may arise due to
intervening low mass objects (intracluster MACHOs).  We discuss this
possibility in the next section.

\section{Intracluster Machos}
\label{sec:machos}

The requirement for high enough a magnification in order to identify a
mismatch in the luminosity-distance relation is essential only if we
like to remove the mass sheet degeneracy. However, even if a SN is
observed at radii larger than $\theta_{\rm max}(\amin)$, it is still
very useful as a point-like probe for 
the dense environment of the cluster. The spectral line
identification, and the family of LCSs, allow for an identification of
many possible modulations of the light curve.
The main source for such modulations are compact objects in the
intracluster medium which induce microlensing events on top of the
cluster magnification.

The Einstein radius of a point mass ${\rm M}$ is
\begin{equation}
  r_{\rm E} = \left(\frac{4G{\rm M}}{c^2}\,D_{\rm eff}\right)^{1/2}\;.
\label{eq:1}
\end{equation}
$D_{\rm eff}=D_{\rm L}D_{\rm LS}D^{-1}_{\rm S}$ is the effective
lensing distance, with $D_{\rm L,S,LS}$ the angular diameter
distance to the lens, the source, and between lens and source, respectively.
Inserting numbers,
\begin{equation}
  r_{\rm E} = 4.3\times10^{16}\,{\rm cm}\,
  \left(\frac{{\rm M}}{{\rm M}_\odot}\right)^{1/2}\,
  \left(\frac{D_{\rm eff}}{{\rm Gpc}}\right)^{1/2}\;.
\label{eq:2}
\end{equation}
The time scale for a source crossing the Einstein disk is
\begin{equation}
  \Delta t_{\rm \mu L} = 
  \frac{2r_{\rm E}}{v_{\rm eff}} = 27.3\,{\rm yr}\,
  \left(\frac{{\rm M}}{{\rm M}_\odot}\right)^{1/2}\,
  \left(\frac{D_{\rm eff}}{{\rm Gpc}}\right)^{1/2}\,
  \left(\frac{v_{\rm eff}}{10^3\,{\rm km\,s^{-1}}}\right)^{-1}\;,
\label{eq:3}
\end{equation}
with the effective relative velocity $v_{\rm eff}\simeq
\sqrt{2}\sigma_v$ of microlens and source. The time scale $\Delta
t_{\rm SN}$ of a SN Ia at redshift $z_{\rm s}\sim1$ is of order 100
days. Even for higher redshifts, this is the
maximum
relevant time-scale (half a year) due to observational constraints.
The microlensing time scale
$\Delta t_{\rm \mu L}$ is therefore compatible with $\Delta t_{\rm
SN}$ only for very small microlens masses, ${\rm M}\sim10^{-4}\,{\rm
M}_\odot$.

The time scale may be shorter due to the expansion of the SN
photosphere.
An estimate for this limit is obtained by requiring 
that the angular size of the SN not exceed 
that of the Einstein radius.
For a rest-frame expansion rate of $v_{\rm SN}\simeq 10^9$ cm
s$^{-1}$, the angular size of the SN photosphere 
at {\it observed\/} time $t$ is $\theta_{\rm
SN}= v_{\rm SN}\,t (1+z_{\rm S})^{-1} D_{\rm S}^{-1}$. This angular size
should be compared to $r_{\rm E}\,D_{\rm L}^{-1}$ to give
\begin{equation}
\Delta t_{\rm lim}= {r_{\rm E}(1+z_{\rm S}) \over v_{\rm SN} }\, {\ds
\over \dl}\, .
\label{eq:dt_lim}
\end{equation}
For $\ds  \dl^{-1} \simeq 10$, ${\rm M}\sim 10^{-4}{\rm M}_\odot$, and
$z_{\rm S}= 1-2$, 
$\Delta t_{\rm lim}$ amounts
to $\sim 30$ days. For smaller masses, or later observational
times, a microlensing event may still be detected, but the typical
microlensing amplification is suppressed, and eventually disappears
(\cf Schneider \& Wagoner 1987).
The SN time scale is thus
$\Delta t_{\rm SN}=\min(100\,{\rm d},\Delta t_{\rm lim})$. 

\subsection{Point mass in a cluster}

An isolated point mass embedded in a SIS cluster is the typical case
of a Chang-Refsdal lens (Chang \& Refsdal 1979). Shear and convergence
of the cluster at the position of the microlens increase the cross
section of the microlens, which we here define to be the area enclosed
by the critical curve of the microlens. While the cross section (in
units of an area rather than a solid angle) of an isolated microlens
is $\tilde\sigma_0=\pi\,r_{\rm E}^2$, the corresponding cross section
of the Chang-Refsdal lens is $\tilde\sigcr (\theta)\simeq
A(\theta)\tilde\sigma_0$, to excellent accuracy. $A(\theta)$ is the
magnification of the cluster at the position $\theta$ of the
microlens, cf.\ eq.~(2-1).

\begin{figure}[t!]
\vskip-2.0truecm
\hskip3.6truecm {\epsfxsize=3.7 in \epsfbox{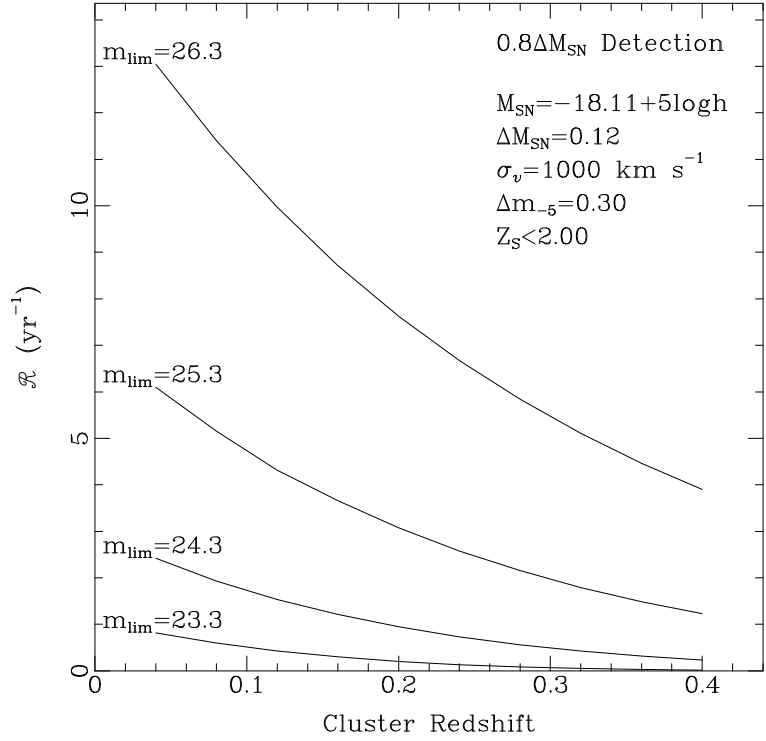}}
\vskip-1.0truecm
\caption{\baselineskip 0.4cm{\trm The expected detection rate of SNe
  Ia in the background of a lensing cluster as function of the lens
  redshift. Minimal magnification of $\amin=1.1$ is assumed
  corresponding to $\theta\simeq 3.5-4.5$
  arcmin.\ in the plotted range.}}
\label{fig:rate_0.8sig}
\end{figure}

Let $n(\theta)$ be the surface 
number density of microlenses at radius $\theta$ in the cluster. We
assume that $n(\theta)$ is a fixed fraction $f$ of the cluster's
surface mass density $\Sigma(\theta)$,
\begin{equation}
  n(\theta) = \frac{f\,\Sigma(\theta)}{\rm M} =
  \frac{f\,\Sigma_{\rm cr}\,\theta_{\rm E}}{2\theta\,{\rm M}}\;,
\label{eq:4}
\end{equation}
where we have used the critical surface mass density
\begin{equation}
  \Sigma_{\rm cr} = \frac{c^2}{4\pi G}\,D_{\rm eff}^{-1}
\label{eq:5}
\end{equation}
and the (angular) Einstein radius of the cluster
(eq. (\ref{eq:a_sis})).  The average number of microlenses straddled
by a macroimage at position $\theta$ is
\begin{equation}
  N(\theta) = n(\theta)\,\tilde\sigcr(\theta)\;.
\label{eq:7}
\end{equation}
Inserting $\tilde\sigma(\theta)=A(\theta)\pi r_{\rm E}^2$,
and using
eqs.~(\ref{eq:1}) and (\ref{eq:5}), eq.~(\ref{eq:7}) becomes
\begin{equation}
  N(\theta) = \frac{f\,A(\theta)\,\theta_{\rm E}}{2\theta}\;.
\label{eq:8}
\end{equation}
This is the equivalent to the
microlensing ``optical depth'' at 
angular distance $\theta$ from the cluster center.  In order to find
the average number of microlenses straddled by {\em any\/} macroimage,
we need to average $N(\theta)$ with the probability distribution for
image positions, $P(\theta)$. Assuming the sources are randomly
distributed in the source plane, we can write (\cf eq.
(\ref{eq:p_of_a}))
\begin{equation}
  P(\theta){\rm d}\theta = (A_{\rm min}-1)^2\,
  \frac{2\,(\theta-\theta_{\rm E})}{\theta_{\rm E}^2}\,
  {\rm d}\theta\;.
\label{eq:9}
\end{equation}
This probability function is still not the probability of observing a
SN at angle $\theta$. The latter is given by the additional magnitude
limit imposed by the observations. For this we need not find a SN with
a luminosity-distance mismatch, but merely a SN that exceeds the
magnitude threshold for identification. The rate of observed SNe is
given by a similar expression to eqs. (\ref{eq:rate},
\ref{eq:rate_d}). We again assume a $\delta$-function for the SN
absolute magnitude distribution, and eliminating the mismatch
constraint we require arbitrarily $\amin=1.1$.
This value is still within the scale where it is
adequate to use the SIS model for the cluster density profile.  Figure
\ref{fig:rate_0.8sig} shows the expected detection rate for SN Ia with
the rest of the parameters identical to those of
fig. \ref{fig:rate}. If $m_{\rm lim}\ge 26.3$ all $\sim10$ expected SN
behind nearby clusters can be used for the macho detection, though
only a small fraction of them ($\sim 1:5$) are suitable for a mismatch
detection.

We do not know {\em a priori\/} where an individual SN will show up in
the cluster background. Based on the probability function (\ref{eq:9})
we can however predict the average optical depth. Combining
eqs.~(\ref{eq:8}) and (\ref{eq:9}), we find
\begin{equation}
  \langle N\rangle = 
  \int_{\theta_{\rm E}}^{\theta_{\rm max}}
  {\rm d}\theta\,P(\theta)\,N(\theta) =
  f\,(A_{\rm min}-1)\;.
\label{eq:10}
\end{equation}
Note that $\langle N\rangle$ does not depend on the mass of the
individual microlenses. The probability for a microlensing event is
therefore independent of the microlens mass, but the microlensing time
scale selects for masses of order $10^{-7}-10^{-4}\,{\rm M}_\odot$.

\begin{figure}[t!]
\vskip-2.0truecm
\hskip3.6truecm {\epsfxsize=3.7 in \epsfbox{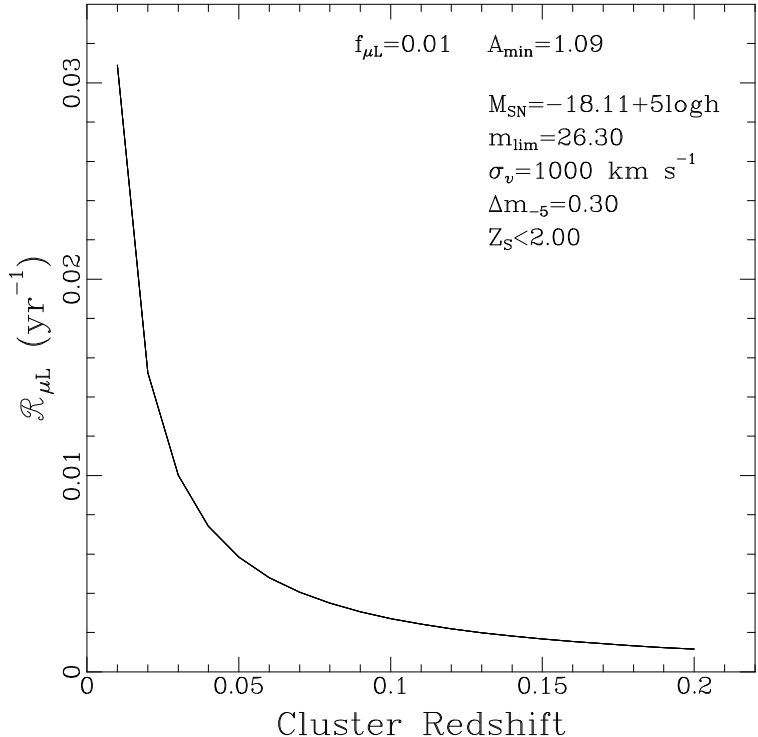}}
\vskip-1.0truecm
\caption{\baselineskip 0.4cm{\trm The expected detection rate of
  microlensing events in the background of a lensing cluster as a
  function of the lens redshift.  We assumed $f=0.01$, $\sigma_v=1000
  \kms$, 
  and $\Delta t_{\rm SN}$ as dictated by the SN expansion rate ($\Delta
  t_{\rm lim}$, Eq. \ref{eq:dt_lim}).
  The magnitude limit for the SN identification is assumed
  to be $26.3$ (see \ref{subsec:rate}), and $\amin=1.09$.  
  The relevant MACHO mass range is $(10^{-4}-10^{-7})\,{\rm M}_\odot$.
  }}
\label{fig:rate_ml}
\end{figure}

In order to estimate the total number per year of microlensing events
that we expect for a given cluster, the probability to observe a SN
should be folded in. The rate is then given by (\cf
eq. \ref{eq:rate_d})
\begin{equation}
  {\cal R}_{\rm \mu L} \simeq
  {f\,(A_{\rm min}-1) \over 100}
  \int_{z_{\rm L}}^\infty {\rm d}z {{\rm d}R(z) \over {\rm d}z} R^2(z) \,
  {S_{\rm SN}(z) \over (1+z) }
  {\Delta t_{\rm SN}(z) \over \Delta t_{\rm \mu L}(z) }
  \times {\rm min}\left[\sigma_{\rm sis}(A')\,,
  \sigma_{\rm sis}(\amin)\right]\,{\rm yr}^{-1}.
\label{eq:rate_ml}
\end{equation}
Figure \ref{fig:rate_ml} shows the expected number of microlensing
events behind clusters of various redshifts. We assumed $f=0.01$
and $\sigma_v=1000 \kms$.
The magnitude limit for the SN identification is
assumed to be $26.3$ (see \ref{subsec:rate}), and $\amin=1.09$ which
corresponds to $\theta \simeq 3.5-4.5$ arcmin.\ in the plotted
redshift range. 
This rate is relevant for the MACHO 
mass range ($10^{-4}-10^{-7}){\rm M}_\odot$.

\begin{figure}[ht]
\hskip2.0truecm
{\epsfxsize=0.7\textwidth\epsffile{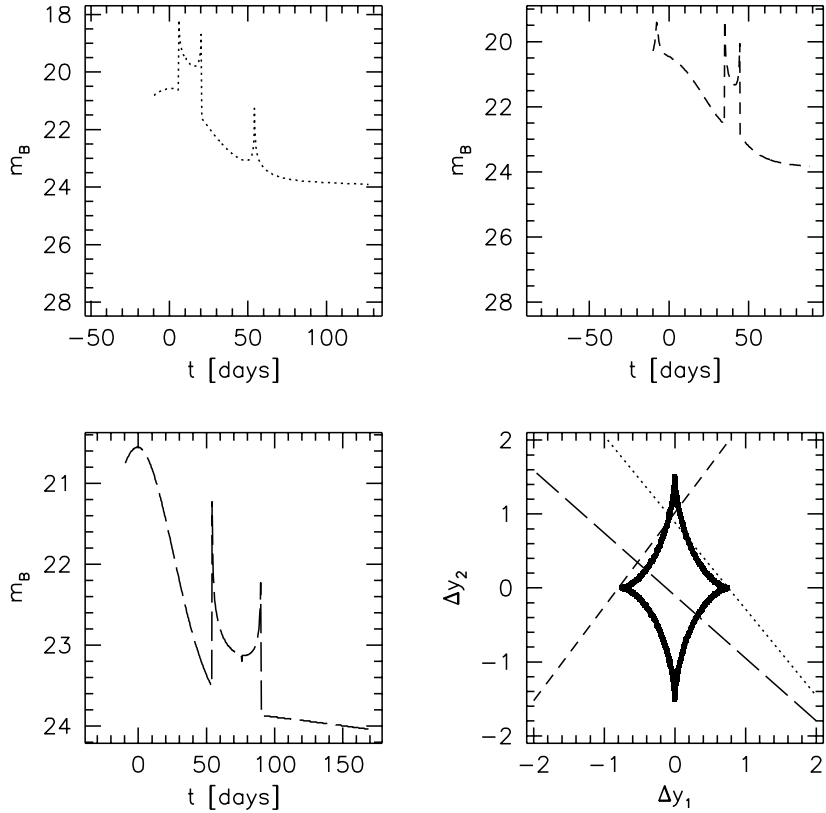}}
\caption{\baselineskip 0.4cm{\trm Examples for three microlensed SN Ia
  light curves. The SNe Ia are assumed at redshift $z_{\rm S}=0.8$,
  the cluster at redshift $z_{\rm L}=0.2$. The source tracks relative
  to the microlens are chosen randomly.  The microlens mass is ${\rm
  M}=10^{-4}\,{\rm M}_\odot$, and the effective relative velocity is
  $v_{\rm eff} = \surd 2\,10^3\,\kms$.  The cluster magnification at
  the position of the microlens was chosen to be 4. $\Delta y_{1,2}$
  are in units of $r_{\rm E}$, projected on the source sphere.
  The lower right panel depicts the caustic curve of the microlens and
  the tracks of the SN relative to the microlens.}}
\label{fig:micro}
\end{figure}

If a microlensing event leaves its signature on the LCS, what would
this signature look like?  Figure \ref{fig:micro} shows examples of
three SN Ia LC's, 
modified by microlensing due to $10^{-4}{\rm M}_\odot$ intracluster
MACHOs.  We modeled the LCS by $-M(d)=(a\,\exp[-(d/\sigma_d)^2] +
s)\times[1+b(d/d_t)]$ with $d$ in days, and the parameters $a=3$,
$\sigma_d=20$, $s=17.2$, $b=-1/63$, and $d_t=50$. 
This functional form, and the 
parameters, provide a curve very similar to the template RPK2 used for
SN 91T [$d\in(-10,150)$ days]. For the microlensing events, we use
$z_{\rm L} = 0.2$, $z_{\rm S}=0.8$, and $\sigma_v=10^3\kms$ for the
SIS cluster profile.  
We assume a SN angular size smaller than the MACHO Einstein radius
throughout the event duration.
The lower right panel of the figure depicts the caustic curve of the
microlens projected onto the source sphere and the 
tracks of the SN relative to the microlens.
The figure nicely demonstrates that it should
not be too hard a task to separate the microlensing event signature
from the LCS of the SN.  The LCSs of the SNe are differently modeled
in different bands. The microlensing event on the other hand appears
identical in all bands thanks to its achromatism. This dichotomy in
nature makes the deconvolution of the two easy to accomplish.

\section{Summary and Conclusions}
\label{sec:conc}

We have proposed to look for SN type Ia behind nearby rich clusters.
The expected rate of detectable SN behind such clusters is of the
order of
$~10$ SN yr$^{-1}$ per cluster, about 20\% of which can be
used alone as tools to remove the mass sheet degeneracy. The rest of
the detected SN can be combined together for the same purpose.  Since
the magnification scales like $\theta^{-1}$ and the number of sources
scales like $\sim \theta^2$ (neglecting magnification bias), we expect
the signal-to-noise ratio to be constant.  Statistical use of the SNe
on large radii yields a similar signal-to-noise ratio as for the SNe
that exceeded the mismatch threshold.

Any of the detected SN can serve as a point source with a typical
light curve shape. Any modulation of this light curve shape, or lack
thereof, can put limits on the mass fraction of the cluster in the
form of $10^{-7}<{\rm M}/{\rm M}_\odot<10^{-4}$ MACHOs.

This proposed survey shifts the focus from high redshift clusters as
targets for high redshift SNe to low redshift clusters as foreground
lenses for the same, and farther SNe.  At the same time, the nearby
observed cluster enhances the probability of finding serendipitous SNe
in it.  The use of the cluster SNe as point sources for MACHO search
is somewhat less efficient because of the relatively small effective
lensing distance $D_{\rm eff}$, and the unknown position of the source
within the cluster.  The monitoring procedure of current observations
will have to be modified, namely to allow for longer gaps between
successive observations of the same cluster, but more observations in
total during the year. The proposal demands a higher magnitude limit
than the values that are currently used, both for the identification
and for the follow-up. These limits, however, are not out of
reach. 

Let us list again all the choices we have made to result in
an underestimate of the mismatch SNe rate.

\begin{enumerate}

\item We do not take into account the fainter image when multiple
images appear on the image plane.
\item We took the specific SN Ia rate at $z=0.4$ 
and used it as the representative value for higher redshifts despite
the prediction of increasing rate with redshift.
\item We choose a $\delta$-function to represent the probability
distribution of the SN absolute blue magnitude. The $\delta$-function
is centered on the mean, even though the median of the distribution
seems to be on the bright side. 
On the other hand, note we have ignored a possible Malmquist bias in the
Hamuy \etal (1996a) sample, a bias that tends to slightly shift the average
towards brighter magnitudes.

\item We considered only SNe that exceed the detection (magnitude
limit) threshold $5$ days (rest-frame) prior to peak luminosity. A
weaker requirement (\ie only $0.2$ magnitudes below peak) may suffice
if a denser time sampling is devised, or if the SN is intrinsically
more luminous than average (and therefore rises and declines more
slowly).
\item We require a mismatch of twice the natural scatter in the
LCS-luminosity relationship. A more modest requirement may be
sufficient to establish limits on the surface mass density along the
line-of-sight.
\item We only integrate in redshift up to $z=2$ because of lack of
knowledge of the number density of galaxies beyond this range.  Note
that if we are not willing to extrapolate the galaxy number density
beyond $z\simeq1.5$, but we still like to satisfy the observed surface
number density of galaxies, the number density should rise even more
steeply than assumed beyond $z=1$. If this is the case, more SN Ia
fall under the magnitude limit of the survey, and the rate increases
correspondingly.

\end{enumerate}

It seems that the main challenge for this proposed method is the high
accuracy photometry needed. Two groups are currently conducting a hunt
for high redshift SNe 
(the SN cosmology project based at Lawrence Berkeley National Laboratory
(Perlmutter \etal 1997); the high-z SN search team based in Mt.
Stromolo, Australia (Schmidt \etal 1996)). 
One survey fits this paper proposal if higher magnitude
limits are applied (Abell cluster SN search, University of Washington).
These surveys' objective is generally quite different from the investigation of
galaxy clusters 
(nevertheless, we have just heard that the SN cosmology
project intend to propose a survey in the spirit of this paper).
The expertise they have acquired for determining the
value of the cosmological background parameters should serve them well
in pursuing the nature and distribution of matter within this background.

\acknowledgements 
We thank Saul Perlmutter for helpful comments on the manuscript.
This work was supported in part by the NASA ATP
grant (NAG 5-3061) (TK) and by the Sonderforschungsbereich 375 of the
Deutsche Forschungsgemeinschaft (MB).


\begin{references}

\reference{}
Bartelmann, M. \& Narayan, R. 1995, \apj, {451}, 60
\reference{}
Bartelmann, M., Narayan, R., Seitz, S., \& Schneider, P. 1996, \apj, {464},
115
\reference{}
Bertschinger, E. 1985, \apjs, {58}, 39
\reference{}
Branch, D. \& Tammann, G.A. 1992 \araa, {30}, 359
\reference{}
Branch, D. \& Miller, D.L. 1993, \apjl,{405}, L5
\reference{}
Branch, D., Romanishin, W., \& Baron, E. 1996, \apj,
{465}, 73 
\reference{}
Broadhurst, T., Taylor, A., Peacock, J. 1995, \apj,
{438}, 49
\reference{}
Cappellaro, E., Turatto, M., Benetti, S., Tsvetkov, D.Yu., Bartunov, O.S.,  \&
Makarova, I.N. 1993a, A\&A, {268}, 472
\reference{}
Cappellaro, E., Turatto, M., Benetti, S., Tsvetkov, D.Yu., Bartunov, O.S.,  \&
Makarova, I.N. 1993b, A\&A, {273}, 838
\reference{}
Cappellaro, E., Turatto, M., Tsvetkov, D.Yu., Bartunov, O.S., Pollas, C., Evans, R., \& Hamuy, M. 1997, \aj, {322}, 431
\reference{} 
Carroll, S.M., Press, W.H., \& Turner, E.L., 1992, \araa, 30, 499
\reference{}
Chang K., Refsdal S. 1979, \nat, {282}, 561
\reference{}
Ellis, R.S., Colless, M., Broadhurst, T., Heyl, J., \& Glazebrook, K. 1996, 
\mnras, {280}, 235
\reference{}
Falco, E.E., Gorenstein, M.V., Shapiro, I.I. 1985, \apj, {289}, L1
\reference{}
Filmore, J.A., \& Goldreich, P. 1984, \apj, {281}, 1
\reference{}
Frieman, J.A. 1996, preprint, astro-ph/9608068.
\reference{}
Goobar, A. \& Perlmutter, S. 1995, \apj, {450}, 14
\reference{}
Gunn J., \& Gott J.R. 1972, \apj, {176}, 1
\reference{}
Hamuy, M., Phillips, M.M., Maza, J., Suntzeff,
N.B., Schommer, R.A., \& Avile\'s, R. 1995, \aj, {109}, 1
\reference{}
Hamuy, M., Phillips, M.M., Schommer, R.A., Suntzeff,
N.B., Maza, J., \& Avile\'s, R. 1996a, \aj, {112}, 2391
\reference{}
Hamuy, M. \etal 1996b, \aj, {112}, 2408
\reference{}
Hoffman, Y. 1988, \apj, {328}, 489
\reference{}
Kaiser, N. 1995, \apjl, {439}, L1
\reference{}
Kaiser, N., Squires, G. 1993, \apj, {404}, 441
\reference{}
Kim, A., Goodbar, A., \& Perlmutter, S. 1996, \pasp, {108}, 190
\reference{}
Kneib, J.-P., Ellis, R.S., Smail, I., Couch, W.T., Sharples, R.M. 1995,
\apj{471}, 643
\reference{}
Kovner, I. \& Paczy\'nski, B. 1988, \apjl, {335}, L9
\reference{}
Lilly, S.J., Cowie, L.L., \& Gardner, J.P. 1991, \apj,
{369}, 79
\reference{}
Lilly, S.J., Tresse, L., Hammer, F., Campton, D., \& Le F\`evre, O. 1995 \apj, {455}, 108
\reference{}
Loveday, J., Peterson, B.A., Efxtathiou, G., Maddox, S.J., 1992, \apj, {390}, 338
\reference{}
Metcalfe, N., Shanks, T., Fong, R., \& Jones, L.R. 1991,
\mnras, {249}, 498
\reference{}
Metcalfe, N., Shanks, T., Fong, R., \& Roche, N. 1995,
\mnras, {273}, 257
\reference{} 
Muller, R.A., Marvin, H.J., Pennypacker, C.R., Perlmutter, S., Sasseen,
T.P., \&  Smith, C.K. 1992, \apjl, {384}, L9
\reference{} 
Navarro, J.F., Frenk, C.S., White, S.D.M., 1996, ApJ, 462, 563
\reference{}
Pain, R. \etal 1995, \apj, {473}, 356
\reference{}
Perlmutter, S. \etal 1995, \apjl, {440}, L41
\reference{}
Perlmutter, S. \etal 1997, \apj, {483}, 565
\reference{}
Phillips, M. 1993, \apjl, {413}, L105
\reference{}
Riess, A.G., Press, W.H., Kirshner, R.P. 1995, \apjl, {438}, L17
\reference{}
Riess, A.G., Press, W.H., Kirshner, R.P. 1996, \apj, {473}, 88
\reference{}
Schmidt, B.P. \etal 1996, Bull. Am. Ast. Soc., 189, 108.05
\reference{}
Schneider, P. \& Wagoner, R.V. 1987, \apj, 314, 154
\reference{}
Schneider, P.  \& Seitz, C., 1995, A\&A, 294, 411
\reference{}
Seitz, C. \& Schneider, P. 1995, A\&A, {297}, 287
\reference{}
Seitz, C. \& Schneider, P. 1996, A\&A, 305, 383
\reference{}
Seitz, C. \& Schneider, P. 1997, A\&A, 318, 687
\reference{}
Smail, I., Hogg, D.W., Yan, L., \& Cohen, J.G. 1995 \apjl, {449}, L105
\reference{}
Soucail, G. \& Fort, B. 1990, A\&A, 243, 23
\reference{}
Squires, G. \& Kaiser, N. 1996, \apj, {473}, 65
\reference{}
Turatto, M., Cappellaro, E., \& Benetti, S. 1994, \aj, {108}, 202
\reference{}
Tyson, J.A. 1988, \aj, {96}, 1
\reference{}
Tyson, J.A. \& Fischer, P. 1995, \apjl, {446}, L55
\reference{}
van den Bergh, S. \& McClure, R.D., 1994, \apj, {425}, 205
\reference{}
Vaughan, T.E., Branch, D., Miller, D.L., \& Perlmutter,
S. 1995, \apj, {439}, 558
\reference{}
Wambsganss, J., Cen, R., Xu, G., Ostriker, J.P. 1997, \apj, {475}, L81

\end{references}
\end{document}